\documentclass[conference]{IEEEtran}
\IEEEoverridecommandlockouts
\usepackage{cite}
\usepackage{amsmath,amssymb,amsfonts}
\usepackage{graphicx}
\usepackage{textcomp}
\usepackage{xcolor}
\usepackage{microtype}
\usepackage{graphicx}
\usepackage{subfigure}
\usepackage{booktabs} 
\usepackage{multirow}
\usepackage{xcolor}
\usepackage{amsmath}
\usepackage{amssymb}
\usepackage{algpseudocode}
\usepackage{algorithm}
\usepackage{forloop}

\algnewcommand\And{\textbf{and}}
\algnewcommand\Or{\textbf{or}}
\DeclareMathOperator*{\argmax}{arg\,max}
\def\BibTeX{{\rm B\kern-.05em{\sc i\kern-.025em b}\kern-.08em
    T\kern-.1667em\lower.7ex\hbox{E}\kern-.125emX}}
\usepackage{pifont}
\begin{document}

\title{Metagenome2Vec: Building Contextualized Representations for Scalable Metagenome Analysis
}

\author{\IEEEauthorblockN{Sathyanarayanan N. Aakur\thanks{Corresponding Author: SN Aakur, saakurn@okstate.edu}, Vineela Indla, Vennela Indla, Sai Narayanan, \\Arunkumar Bagavathi, Vishalini Laguduva Ramnath, Akhilesh Ramachandran}
\IEEEauthorblockA{\textit{Oklahoma State University}\\
Stillwater, OK}
}

\maketitle

\begin{abstract}
Advances in next-generation metagenome sequencing have the potential to revolutionize the point-of-care diagnosis of novel pathogen infections, which could help prevent potential widespread transmission of diseases.
Given the high volume of metagenome sequences, there is a need for \textit{scalable} frameworks to analyze and segment metagenome sequences from clinical samples, which can be highly imbalanced.
There is an increased need for learning robust representations from metagenome reads since pathogens within a family can have highly similar genome structures (some more than $90\%$) and hence enable the  segmentation and identification of novel pathogen sequences with limited \textit{labeled} data. 
In this work, we propose \emph{Metagenome2Vec} - a contextualized representation that captures the global structural properties inherent in metagenome data and local contextualized properties through self-supervised representation learning. We show that the learned representations can help detect six (6) related pathogens from clinical samples with less than $100$ labeled sequences. Extensive experiments on simulated and clinical metagenome data show that the proposed representation encodes compositional properties that can generalize beyond annotations to segment novel pathogens in an unsupervised setting. 
\end{abstract}

\begin{IEEEkeywords}
Scalable Metagenome Analysis, Deep Representation Learning
\end{IEEEkeywords}

\section{Introduction}
Next-generation sequencing technologies have led to the availability of an abundance of genome data collected from heterogeneous sources at a reduced cost. 
The sheer magnitude, complexity, and diversity of the collected data about biological components, processes, and systems are daunting and require robust, scalable analysis platforms to extract actionable knowledge such as the presence of critical pathogens. 
The emergence of zoonotic diseases from novel pathogens, such as the influenza virus in 1918 and SARS-CoV-2 in 2019 that can jump species barriers and lead to pandemic underscores the need for scalable metagenome analysis. 
Traditional bioinformatics-based approaches have enabled great science; however, they appear to be hitting the wall of efficiency and scalability. They are target specific and sometimes fail to decrypt complex relationships in the data to reveal new discoveries. 
There is a need for \textit{pathogen-agnostic} metagenome analysis frameworks that can help segment and cluster metagenome sequences into smaller and concentrated genome reads that belong to the same pathogen species without prior knowledge of disease etiologies. Such an approach will overcome the multiplexing limitations of traditional methods and help detect novel and emerging pathogens that are not routinely probed for. 

\begin{table}[t]
    \centering
    \caption{\textbf{Genome Similarity.} The many of the pathogens that we consider belong to the same family and hence have highly similar genome sequences. The \textit{exact} overlap percentage is shown; genome pairs extremely high overlap are in bold.}
    
    \resizebox{0.99\columnwidth}{!}{
    \begin{tabular}{|c|c|c|c|c|c|c|}
    \toprule
   \textbf{Pathogen}  & \textbf{H. som.} & \textbf{M. bov.}  & \textbf{M.hae.}  & \textbf{P. mul.} & \textbf{T. pyo.} & \textbf{B. tre.} \\ \toprule
    \textbf{H. somni} & - & 0.797 & 0.876 & \textbf{0.884} & 0.794 &\textbf{ 0.847} \\ \midrule
    \textbf{M. bovis} & 0.797 & - & 0.794 & \textbf{0.801} & \textbf{0.806} & \textbf{0.800} \\ \midrule
    \textbf{M. haemo.} & 0.876 & 0.794 & - & \textbf{0.955} & 0.792 & \textbf{0.945} \\ \midrule
    \textbf{P. multoc.} & \textbf{0.884} & \textbf{0.801} & \textbf{0.955} & - & 0.786 & \textbf{0.847} \\ \midrule
    \textbf{T. pyoge.} & 0.794 & \textbf{0.806} & 0.792 & 0.786 & - & 0.789 \\ \midrule
    \textbf{B. trehal.} & \textbf{0.847} & \textbf{0.800} &\textbf{ 0.945} & \textbf{0.847} & 0.789 & - \\ \midrule
    \end{tabular}
    }
    \label{tab:genome_sim}
\end{table}

Metagenome analysis comes with a set of challenges that we term as the {\textit{small} big data problem}. While metagenomes extracted from clinical samples can have millions of nucleotide sequence reads, many of them belong to the host with a few strands of pathogen sequences interspersed. This results in a long-tail distribution between the different genome sequences that can be present in the sample. This problem becomes more apparent when we consider the similarity between the genome of pathogens that belong to the same family, which can be as high as $95\%$. For example, consider the Bovine Respiratory Disease Complex (BRD). This complex multi-etiologic disease affects cattle worldwide and is one of the leading causes of economic distress in the cattle industry. The chief bacterial pathogens are \textit{Mannheimia haemolytica}, \textit{Pasteurella multocida}, \textit{Bibersteinia trehalosi}, \textit{Histophilus somni}, \textit{Mycoplasma bovis}, and \textit{Trueperella pyogenes}, most of which belong to the \textit{Pasteurellaceae} family. 

A comparison of the genome similarity between the common pathogens in BRDC is shown in Table~\ref{tab:genome_sim}. It can be seen that they are highly similar, with \textit{T. pyogenes} showing more variation. Considering that the average bacterial genome is around $3.7$ million bases, a similarity of $80\%$ between the genomes makes the pathogen detection task extremely hard, particularly considering that metagenome reads from clinical or environmental samples can have high amounts of noise due to observation error~\cite{laver2015assessing}. Note that this turns the pathogen identification problem to a fine-grained detection problem where the goal is to identify metagenome sequences unique to each species to definitively detect their presence. Current approaches to machine learning-based metagenome analysis such as DeepMicrobes~\cite{liang2020deepmicrobes} assume that this challenge is alleviated to a certain extent by employing a \textit{targeted sequencing} approach like 16S rRNA followed by alignment. We, on the other hand, work with raw, un-assembled metagenome sequences and do not assume the presence of any pathogens to offer a more general solution to metagenome analysis. 

\begin{figure}[t]
    \centering
        \includegraphics[width=0.955\columnwidth]{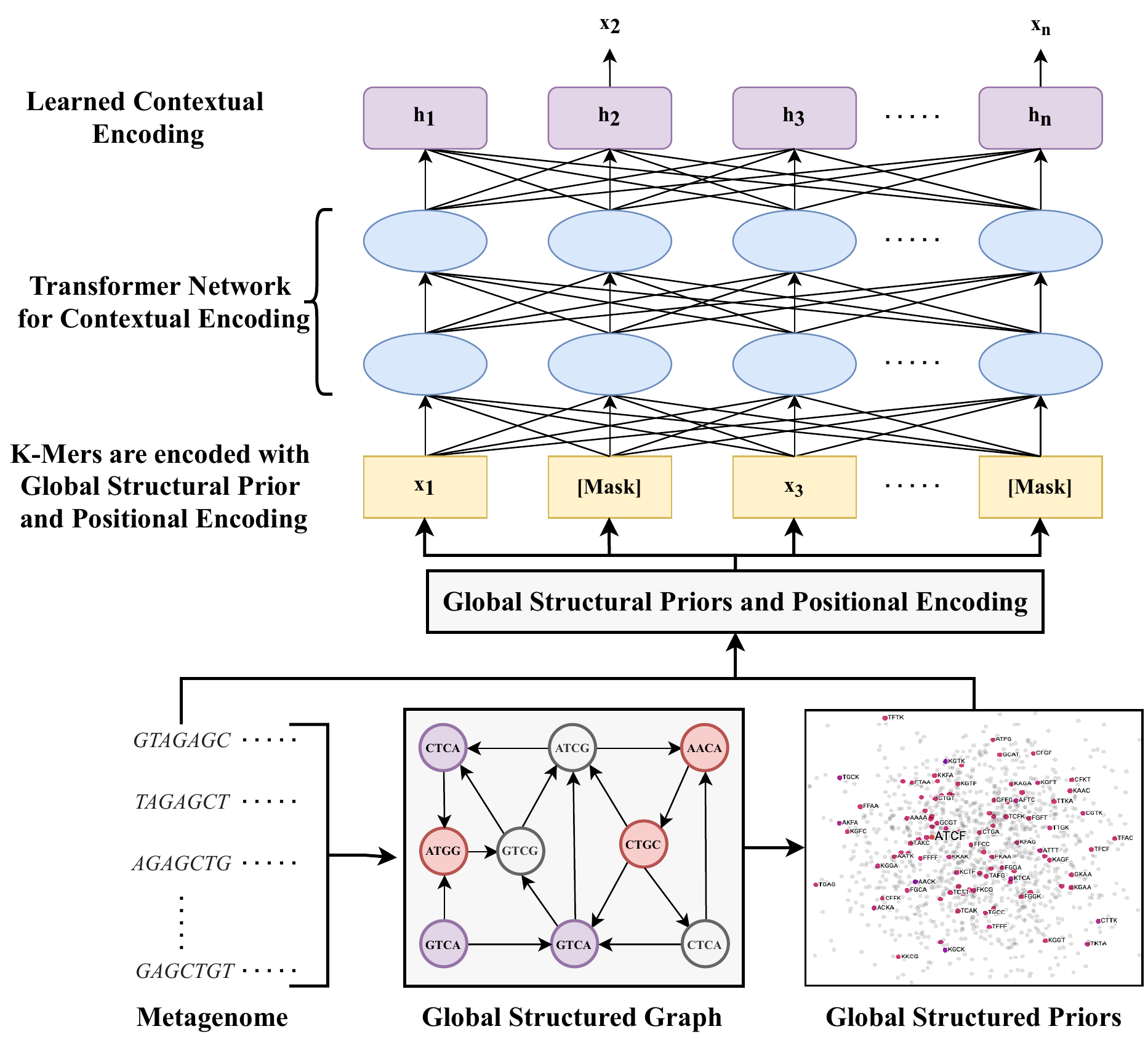}  \\
    \caption{The \textbf{overall approach} for learning robust representations is illustrated. Given an \textit{unlabeled} metagenome sequence, we learn a contextualized representation conditioned upon a global structural prior. The resulting representations can be used for scalable metagenome analysis with minimal labeled data.}
    \label{fig:arch}
\end{figure}

To address this challenging task, we aim to leverage advances in deep learning and computational biology to propose a scalable, metagenome analysis framework to learn robust representations called \textit{Metagenome2Vec}. The goal is to learn robust representations from \textit{unlabeled} metagenome data that can capture fine-grained differences between highly similar pathogen samples. The overall approach is shown in Figure~\ref{fig:arch}. We explicitly capture global structural properties of metagenome sequences and contextualize them with local, sequence-level information to learn robust representations for scalable analysis. We use Bovine Respiratory Disease (BRD) as a model to evaluate \textit{Metagenome2Vec} and show, through extensive experiments, that the resulting representations capture interesting, compositional properties that enable both targeted and generalized pathogen detection. The results show that neural network-based approaches have a great potential to augment the current bioinformatics pipelines such as Kraken2~\cite{wood2019improved} that can be restricted to a given search space. 

\textbf{Contributions:} The contributions of this paper are four-fold: (i) we introduce a novel representation called \textit{Metagenome2Vec} that captures both structural (using graph representations) and contextual properties (using attention-based sequence modeling) from large unlabeled, metagenome sequences, (ii) show that the learned representations can be used for targeted pathogen detection with a high degree of precision with highly limited, unbalanced data with a long-tail distribution, (iii) demonstrate that the learned representations can be used for unsupervised segmentation of clinical metagenome sequences with \textit{unseen} and \textit{unknown} pathogen infections for generalized pathogen detection, and (iv) demonstrate that the learned representations can generalize to completely unrelated pathogens across diseases and species for large-scale metagenome analysis.

\section{Related Work}\label{sec:related}

\textbf{Machine Learning for Metagenomics}: The traditional approaches like BLAST~\cite{altschul1990basic}, Kraken~\cite{wood2014kraken}, Kraken2~\cite{wood2019improved}, and Centrifuge~\cite{kim2016centrifuge} are taxonomy dependent and follow exhaustive search on millions of metagenome sequence reads with a database of target sequence to identify pathogens. Such methods do not scale for large metagenome samples and often do not generalize for novel pathogens since they look for sequence-level matches within a specific search space.
Analysis of DNA sequences collected directly from environment samples offer several opportunities and challenges to apply ML research~\cite{ching2018opportunities} for taxonomy agnostic metagenomics. Deep learning methods in particular are apposite to metagenome analysis due to the availability of large volume of heterogeneous and complex data~\cite{fioravanti2018phylogenetic}. Such taxonomy-agnostic approaches have been utilized to study metagenome classification~\cite{busia2019deep,liang2020deepmicrobes,bartoszewicz2020deepac} and bacteria detection~\cite{fiannaca2018deep}. 

\textbf{Representation Learning for Unstructured Data}: Combining feature representations of unstructured data (like text, images and genome sequences) with features extracted from graphs proved beneficial in multiple applications like rumor detection~\cite{yuan2019jointly}, social community detection~\cite{wang2020integrating}, disease identification in medical images~\cite{wang2020weighted}, and metagenome classification~\cite{Indla2021Sim2RealFM}. Learning low-dimensional feature representations from metagenome sequences are key aspects of existing deep learning algorithms for metagenome classification. Some of the ideas proposed in the literature for metagenome representation learning include: capturing simple nucleotide representations with reverse complement CNNs and LSTMs~\cite{bartoszewicz2020deepac}, learning features from 1-D or 2-D images obtained by clustering metagenome sequences~\cite{nguyen2017deep}, depth-wise separable convolutions to predict taxonomy of metagenome sequences~\cite{busia2019deep}, predicting taxonomy of sequences by learning representations with bidirectional LSTMs with \emph{k}-mer embedding and self attention mechanism~\cite{liang2020deepmicrobes}, and learning metagenome representations to predict the taxonomy~\cite{rojas2019genet}. 

\begin{figure*}
    \centering
    \begin{tabular}{ccc}
         \includegraphics[width=0.29\textwidth]{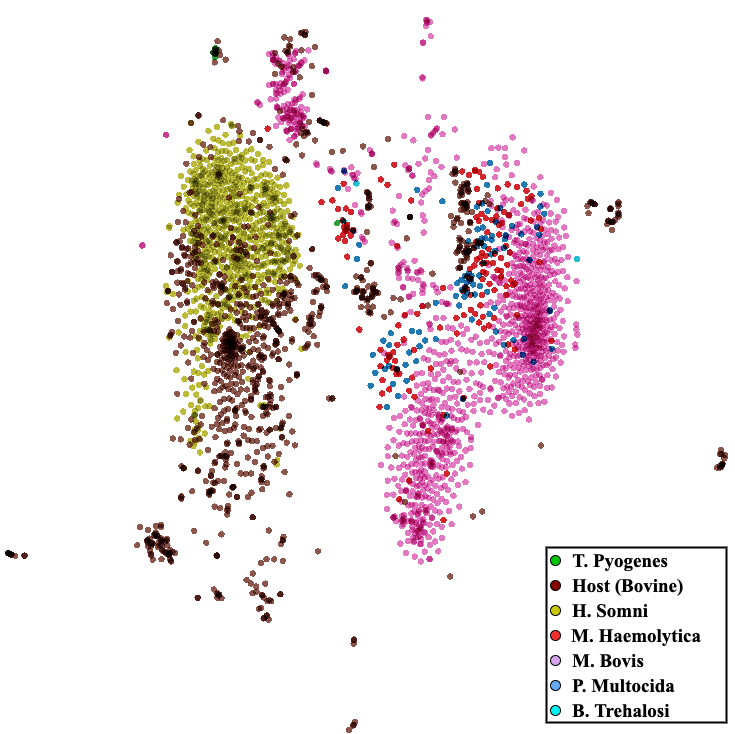} & 
        \includegraphics[width=0.29\textwidth]{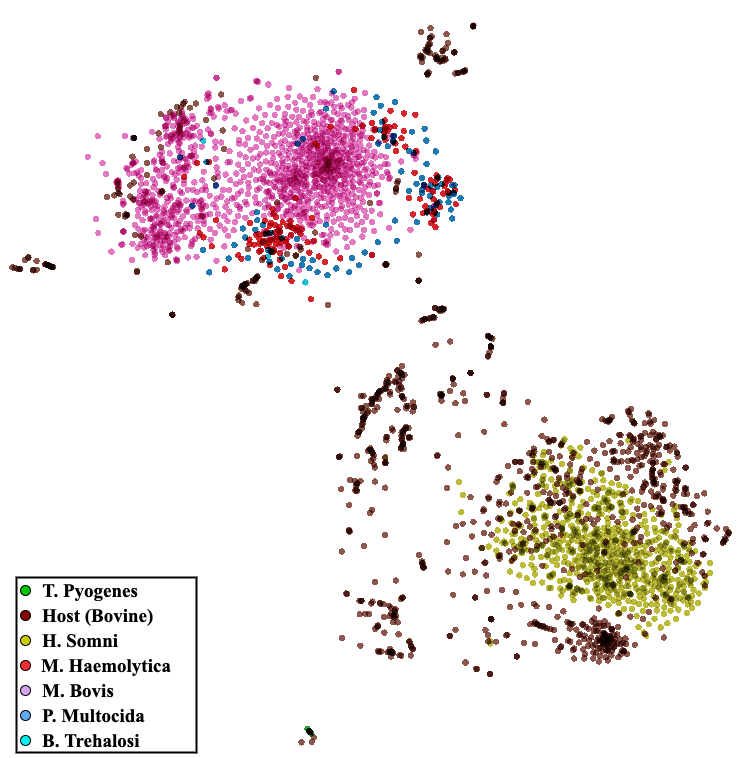}
         \includegraphics[width=0.29\textwidth]{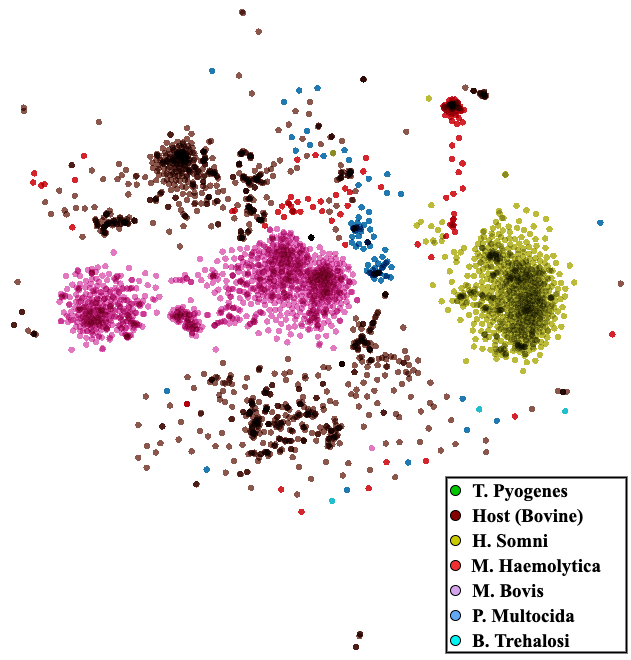}\\
         (a) & (b) & (c)\\
    \end{tabular}
    \caption{\textbf{Visualization} of the features from (a) global structural priors, (b) contextualized sequential representations and (c) \textit{Metagenome2Vec} representations that explicitly code structural and contextual representations. \textit{Metagenome2Vec} provides a much cleaner segmentation between classes.}
    \label{fig:tSNE}
\end{figure*}

\section{Building Contextualized Representations}
\subsection{Background}\label{sec:prob_form}
First, we introduce some necessary background to understand the \textit{scalable} metagenome analysis problem. A metagenome sample consists of a collection of $r$ nucleotide sequence \textit{reads} $\mathcal{X}$, where each \textit{read} $X_i \in \mathcal{X}_r$ belong to an organism $y_i\in \mathcal{Y}_m$. A metagenome sequence read $X_i = \{x_0, x_1, \ldots x_N\}$ is a sequence of $N$ nucleotide bases drawn from a set $\{A, T, C, G\}$. 
The \textit{scalable} metagenome analysis problem aims to learn a robust $D$-dimensional representation $h_{i}\in\mathbb{R}^{1\times D}$ for a given metagenome sequence read $X_i \in \mathcal{X}_r$ to identify its originating organism $y_i\in \mathcal{Y}_m$, with limited annotations. 
In this work, we consider two problem settings to evaluate the scalability and generalization capacity of metagenome analysis frameworks: (i) \textit{targeted} and (ii) \textit{generalized} metagenome analysis. 
In the \textit{targeted} setting, the evaluation is done in a \textit{supervised} learning setting where a limited set of \textit{labeled} \textit{clinical} metagenome samples are available for training. The clinical samples are labeled sequences with a set of known, closely related pathogens that follow a highly unbalanced distribution. 
In the \textit{generalized} setting, the problem is framed as an unsupervised learning problem, where novel, previously \textit{unseen} pathogens are introduced into the \textit{evaluation data}. 
The former setting evaluates the effectiveness of the learned representations for learning with limited \textit{labeled} data while the latter tests the capability to isolate \textit{novel} pathogen sequences to enable discovery and learning of new classes, \textit{with limited human supervision}. 
Combined, these two evaluation settings allow for robust evaluation of machine learning-based metagenome analysis frameworks in real-world settings. 

\subsection{Overall Approach}\label{sec:overallApproach}
We propose \textit{Metagenome2Vec}, a self-supervised representation model that captures both structural and contextual properties from unlabeled metagenome sequences. 
The overall approach is illustrated in Figure~\ref{fig:arch}. Our approach consists of three main components: (i) the construction of a weighted, directed graph from $\mathcal{X}_r$ sequence reads to capture any structural dependencies within the nucleotide co-occurrences, (ii) learning an intermediate representation to capture the global structural priors from the constructed graph structure, and (iii) use an attention-based formalism to \textit{contextualize} and enhance the global structural representations, with local, sequence-level context. 
As can be seen from Figure~\ref{fig:arch}, the first step is to construct a graph from the metagenome sequences that provide a global structural prior for learning robust representations. 
We draw inspiration from the success of De Bruijn graphs~\cite{lin2016assembly} for genome assembly and construct a \textit{weighted, directed} graph from the metagenome sequences. 
Each node in the graph represents a k-mer~\cite{marccais2018asymptotically}, a subsequence from a genome read of length $k$. Note that this is different from the traditional De Bruijn graphs, which are also directed, but \textit{unweighted} multigraph and contain additional edges linking to auxiliary nodes representing ($k{-}1$)-mers and ($k{+}1$)-mers. The direction of the edge is given by the direction of the co-occurrence of k-mers in a given sequence read. The weight of each edge is a function of the frequency existence of an edge between two k-mers detected in metagenome sequences $\mathcal{X}$. Hence, the strength of each edge reflects the \textit{global} frequency of detection of the connection between two k-mers. The construction of the graph is a \textit{streaming} process, with nodes and edge weights updated on the observation of a genome sequence read. Inspired by ConceptNet~\cite{liu2004conceptnet}, the weights between nodes $x_i$ and $x_j$ are updated based on the function given by 
\begin{equation}
    w_{ij} = \lambda_{max} \sqrt{max(\Psi_{i,j} - 1, 1)} + d(\Psi_{i,j}, \lambda_{min})
    \label{eqn:weight_update}
\end{equation}
where $\Psi(x_i, x_j) {=} \frac{e_{i,j}}{(\Vert e_{i,j} - e^{\prime}_{i,j} \Vert_2)}$ and captures the relative increase in the co-occurrence frequency between the k-mers $x_i$ and $x_j$; $e_{i,j}$ is the current edge weight between $x_i$ and $x_j$, and $e\prime_{i,j}$ is the new weight to be updated; $d(\cdot)$ is a function that negates updates in the edge weights for assertions that are weakly reinforced through rare occurrence and is given by $d(\cdot) {=} min(\Psi_{i,j} {-} \lambda_{min}, \lambda_{min}) {+} \lambda_{min}$; and $\lambda_{min}$ and $\lambda_{max}$ are hyperparameters to control normalization bounds. In our experiments, we set both to $2$. 

\subsection{Global Structural Priors}\label{sec:globalStructure}
Learning a low-dimensional representation of the global structural prior in a metagenome is the next step in our proposed framework. This representation must ideally have the following properties: (i) capture the structural similarity between nodes (k-mers) such that they share a set of ``roles'' within the genome sequence and (ii) capture the ``community'' or neighborhood structure of a k-mer to enable the retrieval of relevant nodes due to any noisy sequence reads, which is not that uncommon in next-generation sequencing approaches~\cite{laver2015assessing}. To construct global structural representations with these properties, we use Node2Vec~\cite{grover2016node2vec} as a mechanism to capture the global structure of the metagenome sequences. The global structural representations are constructed using the Skip-Gram approach~\cite{mikolov2013distributed} to capture network structure by the following objective function to maximize the log-probability of observing a structural k-mer neighborhood $N_S(x_i)$ for a node $x_i$ conditioned on its feature representation, given by $h^g_i$ as follows:
\begin{equation}
    \argmax_{h^g_i} \sum_{x_i\in G_r} log \: Pr(N_S(x_i) | h^g_i)
    \label{eqn:globalObjective}
\end{equation}
where $G_r$ is the weighted, directed graph constructed from metagenome sequences $\mathcal{X}$ and $x_i$ is a node corresponding to a unique k-mer. Note that since we construct the graph for the entire metagenome and the edge weights are aggregated using the function from Equation~\ref{eqn:weight_update} and this optimization is done over each node \textit{within} each sequence and averaged across the entire metagenome. The neighborhood $N_S$ is generated by a biased random walk whose transition probabilities are a function of the weight between k-mers (nodes). Hence the neighborhood of each node (k-mer) is a possible pathogen-specific genome sequence that can be found within an overall metagenome infected by the pathogen. The normalization of the edge weights (from Equation~\ref{eqn:weight_update}) allows us to balance the transition probabilities across multiple walks and hence develop a more robust structural prior for each k-mer. 

\begin{table*}[ht]
    \centering
    \caption{\textbf{Recognition results.} Performance of different machine learning baselines on the recognition task on curated, unbalanced metagenome data using the Metagenome2Vec representations. Precision (\textit{Prec.}) and Recall (\textit{Rec.}) are reported for each class.}
    \resizebox{0.99\textwidth}{!}{
    \begin{tabular}{|c|c|c|c|c|c|c|c|c|c|c|c|c|c|c|}
    \toprule
    \multirow{2}{*}{Approach} & \multicolumn{2}{|c|}{\textbf{Host}} & \multicolumn{2}{|c|}{\textbf{H. somni}} & \multicolumn{2}{|c|}{\textbf{M. bovis}} & \multicolumn{2}{|c|}{\textbf{M. haemolytica}} & \multicolumn{2}{|c|}{\textbf{T. pyogenes}} & \multicolumn{2}{|c|}{\textbf{P. multocida}} & \multicolumn{2}{|c|}{\textbf{B. trehalosi}} \\
    \cline{2-15}
     & Prec. & Rec. & Prec. & Rec. & Prec. & Rec. & Prec. & Rec. & Prec. & Rec. & Prec. & Rec. & Prec. & Rec.\\
    \toprule
    LR & 0.97 & \textbf{0.99} & 0.80 & 0.68 & 0.86 & 0.94 & 0.65 & 0.61 & \textbf{0.85} & 0.73 & 0.00 & 0.00 & 0.00 & 0.00 \\ 
    SVM & 0.97 & 0.98 & 0.75 & 0.66 & 0.82 & \textbf{0.96} & \textbf{0.67} & 0.34 & 0.67 & 0.13 & 0.00 & 0.00 & 0.00 & 0.00\\ 
    MLP & 0.98 & \textbf{0.99} & 0.82 & \textbf{0.77} & \textbf{0.89} & 0.94 & 0.60 & 0.69 & 0.78 & 0.93 & 0.00 & 0.00 & 0.00  & 0.00 \\ 
    DL  & \textbf{0.99} & 0.97 & \textbf{0.84} & 0.73 & 0.87 & 0.95 & 0.44 & \textbf{0.82} & 0.68 & \textbf{1.00} & \textbf{0.50} & \textbf{0.26} & \textbf{ 0.10} & \textbf{0.33} \\ 
    \midrule
    K-Means & 0.95 & 0.36 & 0.33 & 0.75 & 0.55 & 0.87 & 0.24 & 0.12 & 0.62 & 1.00 & 0.47 & 0.41 & 0.00 & 0.00 \\ 
    \bottomrule
    \end{tabular}
    }
    \label{tab:quant_result}
\end{table*}

\subsection{Contextualized Representations with Attention}\label{sec:pre_training}
The next step in building robust metagenome representations is \textit{contextualization}. We consider a representation to be \textit{contextualized} if the resulting representation can (i) reduce the impact of background or spurious structural or sequential patterns that may occur due to noisy observation and (ii) \textit{integrate} a given k-mer's feature representation with its surrounding context. This representation is analogous to a posterior-weighted representation that highlights areas of interest while suppressing contextually irrelevant features. 
To build such representations, we use the idea of \textit{word embedding} to learn a mapping function $M_c{:} x_i\rightarrow \mathbb{R}^{1\times D}$. However, to ensure that the context is captured effectively we leverage the attention-based transformer~\cite{vaswani2017attention} to learn our contextualized mapping function. 
We initialize the mapping function $M$ with the global structural embedding from Section~\ref{sec:globalStructure} and use the transformer to capture contextual properties in each sequence. 

A transformer is a neural network architecture composed of a series of multi-headed self-attention (MHA) and token-wise feedforward operations. At each level of the network, the representation of each token is updated based on the hidden representation from the lower levels. In our architecture, the k-mers, extracted from each sequence read ($\mathcal{X}_j{=}x_1, x_2, \ldots x_N; \forall X_j {\in} \mathcal{X}_r$), are fed as tokens to the network and the hidden representations of the final layer $H{=}h_1, h_2, \ldots h_D$ represent the learned, contextual representations for each k-mer. The transformer representations are conditioned on the global structural prior by initializing the embedding layer with the mapping function $M$. 
To ensure that these representations further capture the contextual properties from the entire sequence, we allow the model to use bidirectional attention by employing a masking approach, as common in masked language models such as BERT~\cite{devlin2019bert}. Specifically, we randomly \textit{mask} or replace a subset of tokens in the input $X_j$ with a [MASK] token and train the network to predict these missing tokens. Hence the objective function for the network is to minimize the log-likelihood of the masked tokens $X\mathcal{X}^\Gamma_j$ given the observed tokens ($\mathcal{X}^{-\Gamma}_{j}$) and is represented as:
\begin{equation}
    \mathcal{L}(\mathcal{X}^\Gamma_j | \mathcal{X}^{-\Gamma}_{j}) = \frac{1}{T}\sum_{t=1}^{T}log \: p(x^{\Gamma}_{t} | \mathcal{X}^{-\Gamma}_{j};\theta)
    \label{eqn:mlm_objective}
\end{equation}
where each genome sequence is given by $\mathcal{X}_j{=}\mathcal{X}^{-\Gamma}_{j}\bigcup \mathcal{X}^{\Gamma}_{j}$ and $\mathcal{X}^{\Gamma}_{j}{=}\{x^{\Gamma}_1, \ldots x^{\Gamma}_t\}$ is a set of $T$ masked tokens; $\theta$ is the set of learnable parameters in the network. The masks are chosen such that the probability of masking is given by $p(M){=}s^T(1-s)^{N-T}$ where $s$ is the masking ratio. 

The objective function from Equation~\ref{eqn:mlm_objective} is used to train the transformer architecture, and more importantly, the errors are allowed to propagate all the way to the initial mapping function $M$, which is initialized with the global structural priors. The updated mapping function ($M_c$) provides allows us to generate \textit{contextualized} embeddings for each k-mer in a sequence and allow us to provide a balance between the global structural prior from the \textit{metagenome} and the local contextual features within each genome sequence read. The final \textit{Metagenome2Vec} representation a sequence read $h^m_i$ is then constructed by the concatenation of the global representation $h^g_i$ and the contextual representation $h^c_i$ for a given k-mer $x_i$. Hence, $h^m_i{=}[h^g_i;h^c_i]$ where $h^g_i{=}M(x_i)$ and $h^c_i{=}M_c(x_i)$.
Note that this is different from the current use of transformers~\cite{devlin2019bert,radford2019language} where the output of the encoder $H$ is used as the extracted features. 
Our approach explicitly encodes the global structural, and local contextual properties of k-mers within a metagenome is similar in spirit to ELMo~\cite{peters2018deep}, which aims to capture the context-dependent aspects of word meaning using bidirectional LSTMs~\cite{hochreiter1997long}. 
To highlight the importance of such explicit representations, we visualize a subset of clinical metagenome sequences for all three representations - global, contextual and \textit{Metagenome2Vec} using TSNE~\cite{van2008visualizing} in Figure~\ref{fig:tSNE}. As can be seen, the explicit representation of structural and contextual properties allows for cleaner segmentation between sequence reads across species. 
Experimental evaluation (presented in Section~\ref{sec:results}) corroborate the observation from visual inspection.

\subsection{Implementation Details}\label{sec:impl_details}
In our experiments, we use $k=4$ to extract k-mer representations with a stride of $1$ on sequence reads. We use this setting based on genome analysis studies that have shown that tetranucleotide bases from phylogenetically similar species can be very similar between closely related species~\cite{perry2010distinguishing}. 
The resulting structural graph has $1296$ nodes, which forms our vocabulary. 
We use Node2Vec on the global structural graph to obtain structural priors. The number of walks per source node is set to $10$, and the walk length is set to $80$. 
We use a transformer with 4 hidden layers and 8 attention heads with a hidden size of $512$ and a dropout rate of $0.1$. We pre-train for $10$ epochs on two unlabeled metagenome samples with a combined total of $2{,}151{,}436$ sequence reads and a batch size of $16$ and the learning rate schedule from~\cite{vaswani2017attention}. 
The training was done on a server with 2 Titan RTXs and a 64 core AMD CPU and took around $36$ hours to converge. 

\section{Experimental Setup}
\textbf{Data.} To obtain real-world metagenome data, we extracted sequences from $13$ Bovine Respiratory Disease Complex (BRDC) lung specimens From the Oklahoma Animal Disease Diagnostic Laboratory (OADDL). 
DNA sequences are extracted from lung samples using the DNeasy Blood and Tissue Kit (Qiagen, Hilden, Germany). 
Sequencing libraries are prepared from the extracted DNA using the Ligation Sequencing Kit and the Rapid Barcoding Kit. 
Prepared libraries are sequenced using MinION (R9.4 Flow cells), and sequences with an average Q-score of more than $7$ are used in the final genome. 
Pathogens in the metagenome sequence data are identified using the MiFi platform~\cite{espindola2021microbe}\footnote{https://bioinfo.okstate.edu/}. Using a modified version of the bioinformatics pipeline~\cite{stobbe2013probe}, \textit{e-probes}, or unique signatures, for each pathogen were developed to identify and label pathogen specific sequences in the metagenome reads.
Samples from $7$ patients were used for training while sequences from $5$ patients were used for evaluation. 

\textbf{Pathogens used for evaluation.} In this work, we consider the pathogens associated with the Bovine Respiratory Disease Complex (BRD). This complex multi-etiologic disease affects cattle worldwide and is one of the leading causes of economic distress in the cattle industry. 
The chief bacterial pathogens of interest are \textit{Mannheimia haemolytica}, \textit{Pasteurella multocida}, \textit{Bibersteinia trehalosi}, \textit{Histophilus somni}, \textit{Mycoplasma bovis}, and \textit{Trueperella pyogenes}, which mostly belong to the \textit{Pasteurellaceae} family. 
For novel pathogen evaluation (Section~\ref{sec:generalDetection}), we considered a simulated metagenome with the following \textit{viral} and bacterial pathogens: Bovine viral diarrhea virus (BVDV), Bovine parainfluenza virus 3 (BPIV-3), Bovine herpesvirus 1 (BoHV-1), Bovine coronavirus (BCoV) and Bovine respiratory syncytial virus (BRSV). 

\subsection{Metrics and Baselines}
To quantitatively evaluate our approach, we use precision, recall, and the F1-score for each class. We do not use accuracy as a metric since real-life metagenomes can be highly skewed towards host sequences, and it would be possible to obtain high accuracy ($>90\%$) by only predicting the dominant class (i.e., the host). 
We compare against comparable representation learning frameworks for metagenome analysis proposed in literature such as graph-based approaches~\cite{narayanan2020genome}, and a deep learning model termed S2V, an end-to-end sequence-based learning approach based on DeePAC~\cite{bartoszewicz2020deepac}. The S2V baseline is an adaptation of the models used in DeePAC~\cite{bartoszewicz2020deepac} and DeepMicrobes~\cite{liang2020deepmicrobes} for the single-read shotgun sequencing problem setup. For classification, we consider a balanced mix of both traditional machine learning baselines and deep learning baselines. Traditional baselines include Support Vector Machines (SVM), Logistic Regression (LR), and a feed-forward neural network (NN) with two hidden layers with 256 neurons each. The deep learning baselines has 3 hidden layers with 256, 512 and 1024 neurons each with a ReLU activation function. We choose the hyperparameters for each of the baselines using an automated grid search and the best performing models from the validation set were taken for evaluation on the test set.
\begin{table}[t]
    \centering
    \caption{
    \textbf{Comparison with other representations.} Performance evaluation of machine learning baselines using other metagenome representations. Average F1 scores are reported.
    }
    \resizebox{0.99\columnwidth}{!}{
    \begin{tabular}{|c|c|c|c|c|c|c|c|c|}
    \toprule
    \multirow{2}{*}{\textbf{Approach}} & \multicolumn{2}{|c|}{\textbf{Node2Vec}} & \multicolumn{2}{|c|}{\textbf{SPK}} & \multicolumn{2}{|c|}{\textbf{S2V}} & \multicolumn{2}{|c|}{\textbf{MG2V}}\\
    \cline{2-9}
    & \textbf{Host} & \textbf{Path.} & \textbf{Host} & \textbf{Path.} & \textbf{Host} & \textbf{Path.}  & \textbf{Host} & \textbf{Path.} \\
    \toprule
    LR       & 0.824  & 0.044  & 0.857  & 0.128  & - & - & 0.979  & 0.509 \\
    SVM      & 0.806  & 0.071 &  0.863  & 0.113  & - & - & 0.975  & 0.405 \\
    MLP      & 0.849  & 0.097  & 0.864  & 0.080  & -  & - & \textbf{0.985}  & 0.534 \\
    DL       & 0.744  & 0.102  & 0.783  & 0.099  & 0.758 & 0.362 & 0.981  & \textbf{0.631}\\
    \bottomrule
    \end{tabular}
    }
    \label{tab:graph_rep_comp}
\end{table}
\section{Quantitative Evaluation}\label{sec:results}
As described in Section~\ref{sec:prob_form}, we evaluate metagenome analysis frameworks in two different settings. First, we consider a \textit{supervised} setting called targeted pathogen detection, where the goal is to identify a known number of highly related pathogens from clinical metagenome data. Here, a limited number of training samples are available per class for training. Second, we consider an \textit{unsupervised} setting where we evaluate the robustness of the learned features to segment and isolate multiple pathogen sequences from clinical metagenome samples, including \textit{unseen} closely related bacteria sequences and \textit{unseen} virus sequences. 

\subsection{Targeted Pathogen Detection} 
In this evaluation setting, we extract contextualized representations from the metagenome sequences and train various machine learning models to classify each sequence into a class denoting the host (Bovine) or one of the six (6) pathogen classes. The results are summarized in Table~\ref{tab:quant_result}. It can be seen that our \textit{Metagenome2Vec} representations are robust enough to distinguish between highly similar pathogen sequences with smaller and simpler machine learning baselines such as logistic regression. In fact, a simple k-means clustering approach was able to achieve reasonable performance on this benchmark, indicating that the learned representations are relatively segmented in the feature space. Of particular interest is that there is a perfect recall on \textit{T. pyogenes} which is one of the pathogens not in the same family as the others. A deep learning model, trained with weighted cross-entropy, has the better overall performance across all pathogen sequences. With reasonable precision, it can identify \textit{P. multocida} and \textit{B. trehalosi}, which are the least represented classes in the training set with $37$ and $17$ labeled sequences, respectively. 

However, we find that these simple baselines do not always work in this task, as indicated by their performance when using other contemporary feature representations as seen from Table~\ref{tab:graph_rep_comp}. We compare the performance of the same baselines when presented with features (Node2Vec and Shortest Path Kernel~\cite{borgwardt2005shortest}) from GRaDL~\cite{narayanan2020genome}, an approach similar in spirit to ours, which proposes to use graph representations from \textit{individual} genome sequences for pathogen detection. We also compare against the sequence-based feature representation approach S2V, adapted from DeePAC~\cite{bartoszewicz2020deepac} (S2V) on our data. It can be seen that although the genome sequence representations are learned in a similar manner, they do not scale to the long-tail distribution in the limited, unbalanced labeled data. 

\begin{figure*}
    \centering
    \begin{tabular}{ccc}
    \includegraphics[width=0.3\textwidth]{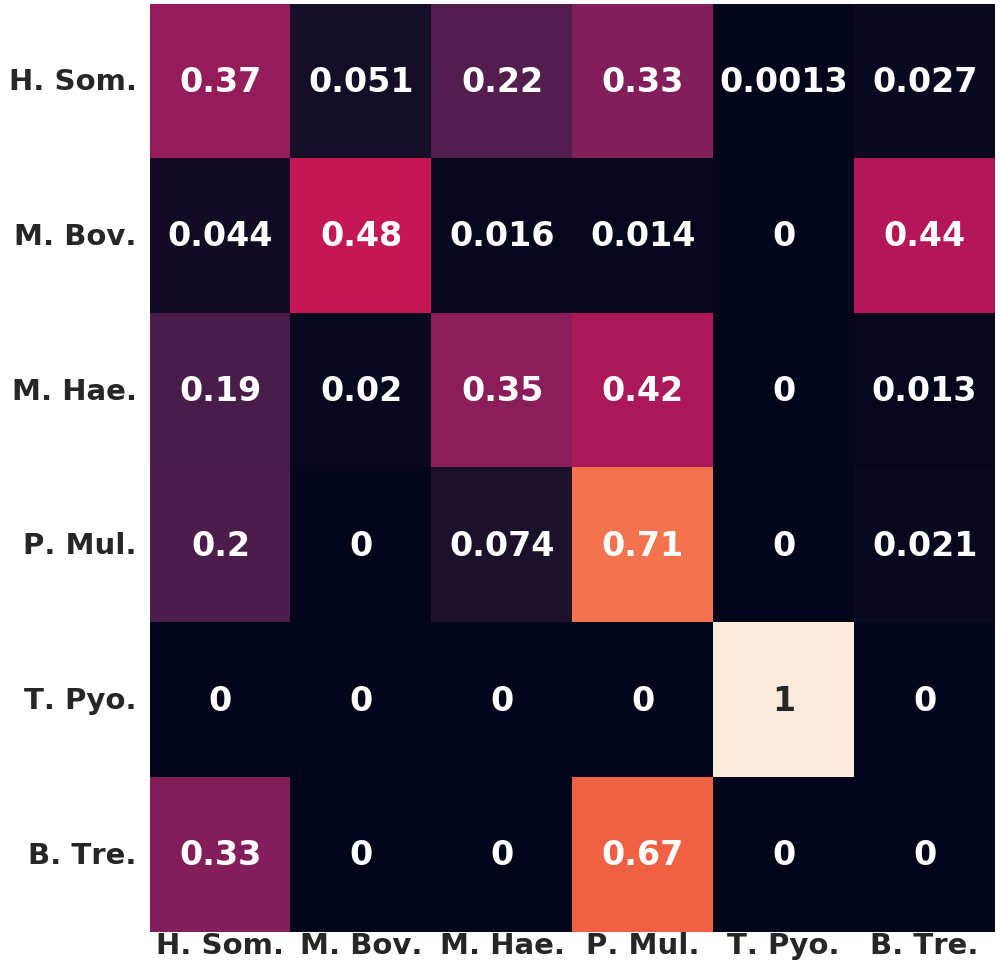} & 
    \includegraphics[width=0.33\textwidth]{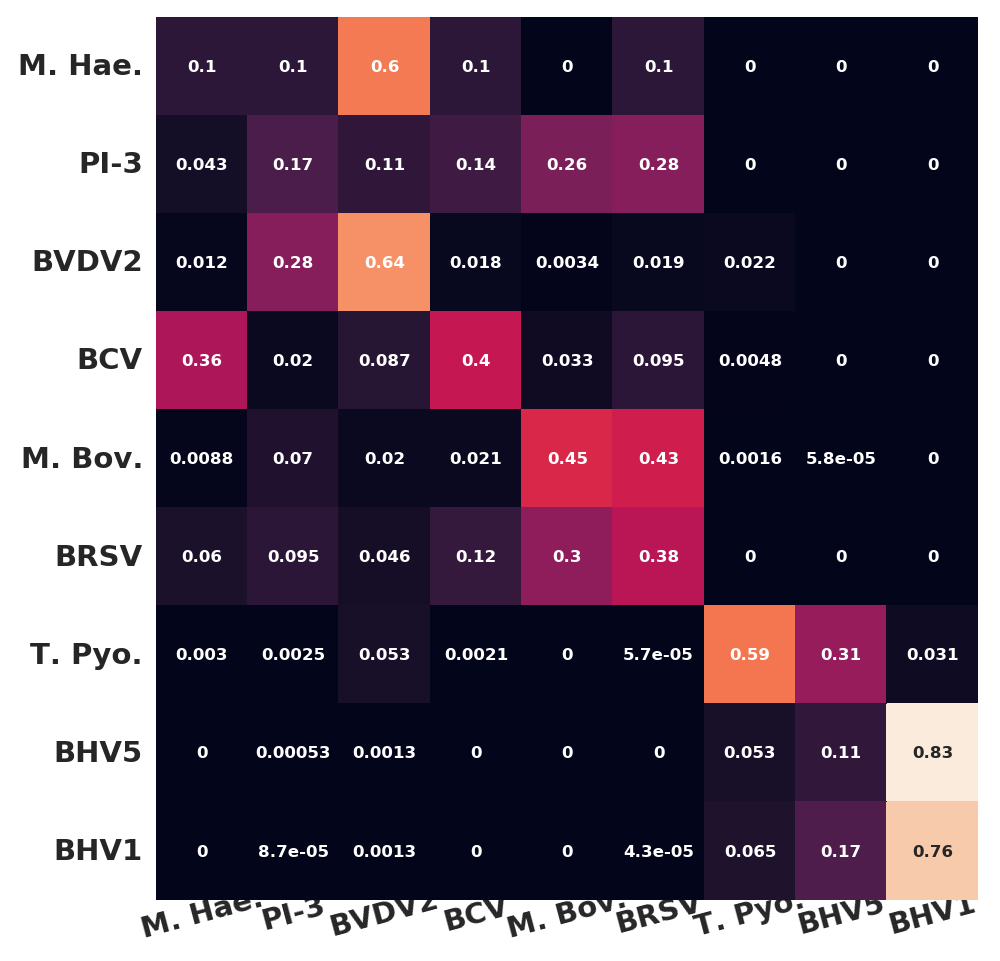} & 
    \includegraphics[width=0.29\textwidth]{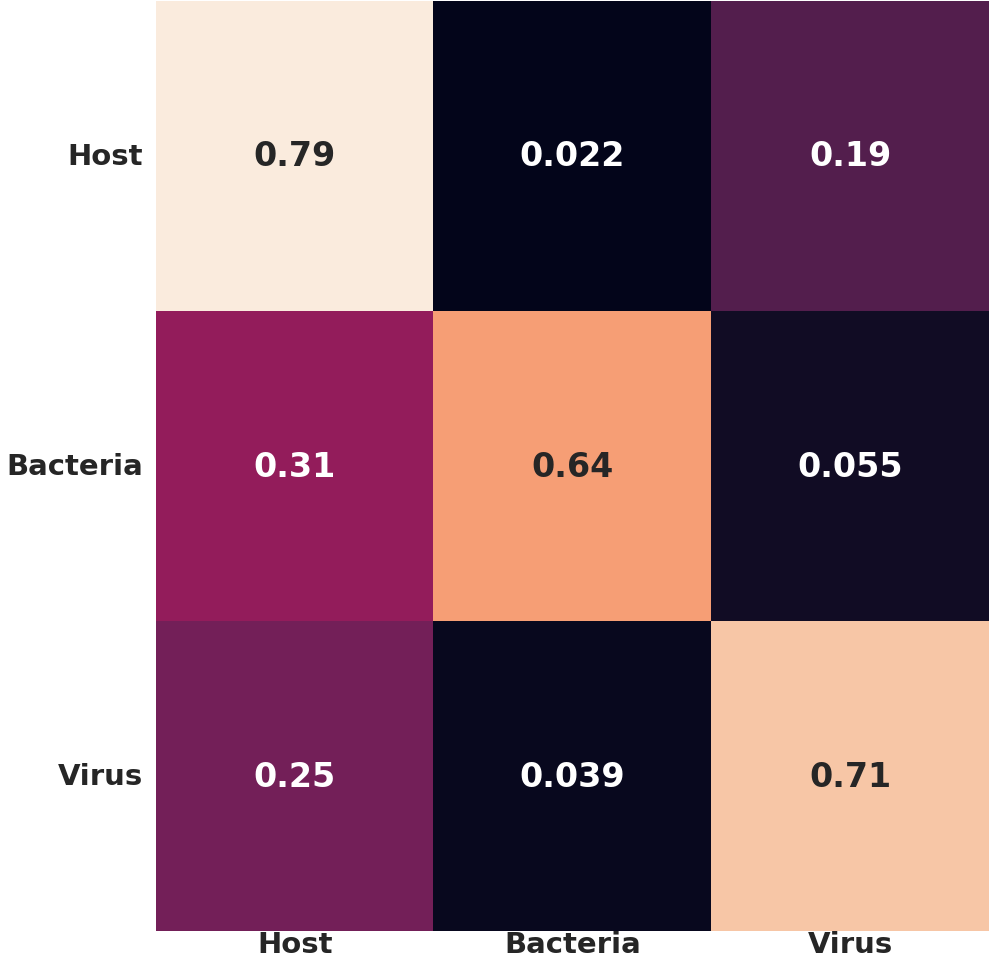} \\
    (a) & (b) & (c)\\
    \end{tabular}
    \caption{\textbf{Feature robustness.} Confusion matrices from unsupervised segmentation of a metagenome containing (a) only seen pathogens and (b) both seen and unseen pathogens, including unseen viruses. (c) shows the confusion matrix for unsupervised segmentation of a metagenome into pathogen groups. }
    \label{fig:pathogen_clustering}
\end{figure*}

\begin{table}[t]
    \centering
    \caption{\textbf{Quality of intermediate representations.} Performance of machine learning baselines with intermediate representations. Average F1 scores are reported.
    }
    \resizebox{0.99\columnwidth}{!}{
     \begin{tabular}{|c|c|c|c|c|c|c|c|c|}
    \toprule
    \multirow{2}{*}{\textbf{Approach}} & \multicolumn{2}{|c|}{\textbf{G-N2V}} & \multicolumn{2}{|c|}{\textbf{C2V}} & \multicolumn{2}{|c|}{\textbf{Encoder}} & \multicolumn{2}{|c|}{\textbf{MG2V}}\\
    \cline{2-9}
    & \textbf{Host} & \textbf{Path.} & \textbf{Host} & \textbf{Path.} & \textbf{Host} & \textbf{Path.} & \textbf{Host} & \textbf{Path.} \\
    \toprule
    LR       & 0.979  & 0.515  & 0.979  & 0.526  & 0.915 & 0.233  &  0.979  & 0.509 \\
    SVM      & 0.975  & 0.525  &  0.979 & 0.536  & 0.904 & 0.171  &  0.975  & 0.405 \\
    MLP      & \textbf{0.985}  & 0.542  & 0.979  & 0.543  & 0.935 & 0.369  &  \textbf{0.985}  & 0.534 \\
    DL       & 0.979  & 0.565  & 0.979  & 0.567  & 0.861  & 0.276   & 0.981  & \textbf{0.631}\\
    \midrule
    K-Means  & 0.426  & 0.225  & 0.471  &  0.245  & 0.358  & 0.144  & 0.441  & 0.368 \\
    \bottomrule
    \end{tabular}
    }
    
    \label{tab:rep_ablation}
\end{table}

\subsection{Generalized Pathogen Detection.} \label{sec:generalDetection}
Given the promising results from \textit{targeted} pathogen detection, we evaluate our approach on the \textit{generalized} pathogen detection task. Here we extract \textit{Metagenome2Vec} representations from a \textit{complete}, unlabeled, metagenome sample and cluster them to segment the metagenome into groups that comprise samples from each pathogen. The cluster labels are mapped to ground-truth labels using the Hungarian method for quantitative analysis. 
This evaluation allows us to identify the capability of building a general-purpose metagenome analysis pipeline that can enable novel pathogen discovery with minimal supervision. 
We use three different settings for generalized pathogen detection. First, we see how well a metagenome sequence containing only previously seen, \textit{unlabeled} pathogens is segmented. The confusion matrix is shown in Figure~\ref{fig:pathogen_clustering}(a). It can be seen that pathogens that are distinct from each other i.e., \textit{P. multocida} and \textit{T. pyogenes}, are isolated pretty well, whereas there is more confusion among highly related pathogens.  
Note that although some part of the genome of these pathogens were observed during pre-training, they were \textit{unlabeled} and hence trained in a \textit{species-agnostic} manner. 

We also evaluate the capability of the representations to generalize to \textit{novel} pathogens, including viruses, that are not part of the pre-training pipeline. We evaluate on a simulated metagenome sample containing one previously \textit{seen} pathogen (\textit{M. haemolytica}) and $9$ \textit{unseen} pathogens. The confusion matrix from this evaluation is shown in Figure~\ref{fig:pathogen_clustering}(b). It can be seen that the \textit{Metagenome2Vec} representations group the metagenome reads very well and in a manner that suggests that the representations implicitly capture the taxonomic-level relationships between the pathogen species. For example, the two related viruses (BHV5 and BHV1), which share $91\%$ of their genome, are confused the most, but not with the more distantly related Bovine Coronavirus (BCV). Similarly, the PI-3 and BRSV viruses belong to the same family \textit{Paramyxoviridae} and the prediction confusion between the two classes is significantly higher. 

Finally, we evaluate the capability of the representations to help segment a metagenome sample into groups from different pathogen types (bacteria vs. virus) to accelerate the targeted search process with established bioinformatics pipelines for fine-grained point-of-care diagnosis. We extract the \textit{Metagenome2Vec} representations and cluster the sequence reads into $3$ groups - host, bacteria and virus and present a visualization in Figure~\ref{fig:pathogen_clustering}(c). It can be seen that the representations are robust to segment out host sequences with high precision (F1-score: 0.76) and help distinguish between bacteria and viral sequences with high fidelity, \textit{without training.} 

\textbf{Comparison with traditional pipelines.} We also compared our approach's performance on generalized pathogen detection with a traditional bioinformatics pipeline using Kraken2~\cite{wood2019improved}. The standard nucleotide database was used for matching metagenome reads, and Kraken2 returned $3131$ unique detections. For a fair comparison, we filter the detected classes to those found within the metagenome sample and add the extraneous detections to a separate class. This setting is similar to our unsupervised segmentation setting, where the set of clusters is fixed, and the metagenome is clustered into its groups containing genome sequences belonging to the same species. Kraken2 obtains an average F1 score of $0.537$, while our unsupervised segmentation achieves an F1-score of $0.409$ when the number of clusters is set to the actual number of pathogens, and $0.624$ when allowed to over-segment. 
Without binning the erroneous detection, the F1-score is $22.87\%$ which is significantly lower than our unsupervised F1-score of $40.9\%$. These are encouraging results considering that our representations are learned from metagenomes where these pathogens, particularly the DNA viruses, were unobserved.

\subsection{Ablation Studies}\label{sec:ablation}
We also systematically evaluate the contributions of the different components of the proposed approach and present each evaluation in detail below. 

\textbf{Intermediate Representations.} We first evaluate the effectiveness of the different intermediate representations that can be obtained from our approach and summarize the results in Table ~\ref{tab:rep_ablation}. We consider four different intermediate representations - only the global structural priors from Section~\ref{sec:globalStructure} (G-N2V), only the contextualized representations from the transformer network's embedding layer from Section~\ref{sec:pre_training} (C2V), the output of the transformer encoder network $H_i$ from Section~\ref{sec:pre_training} (Encoder) and our final \textit{Metagenome2Vec} representation (MG2V). While the other three representations distinguish between host and pathogen, they do not help distinguish among the fine-grained classes within the pathogens as well as MG2V. The structural priors by themselves do well considering that it does not take any pre-training or significant computation time, while the encoder outputs surprisingly do not capture the fine-grained differences among pathogens. 

\begin{table}[t]
    \centering
    \caption{\textbf{Ablation Study:} Evaluation of design choices in the proposed approach such as \textit{k-mers} and structural priors. Precision (\textit{Prec.}), Recall and F1 scores \textit{averaged across classes} are reported. }
    \resizebox{0.99\columnwidth}{!}{
    \begin{tabular}{|c|c|c|c|}
    \toprule
    \textbf{Approach} & \textbf{Prec.} & \textbf{Recall} & \textbf{F1}\\
    \midrule
    \multicolumn{4}{|c|}{\textit{Effect of K-Mers}}\\
    \midrule
    $k=3$ & 0.61 & 0.66 & 0.634\\
    \textbf{$k=4$} & \textbf{0.631} & \textbf{0.723} & \textbf{0.674}\\
    $k=6$ & 0.628 & 0.664 & 0.646\\
    \midrule
    \multicolumn{4}{|c|}{\textit{Effect of Structural Priors}}\\
    \midrule
    without Global Prior  & 0.614 & 0.667 & 0.640 \\
    without Normalized Weights & 0.564 & 0.647 & 0.603 \\
    with Unidirectional Attention  & 0.603 & 0.667 & 0.633 \\
    \midrule
    \multicolumn{4}{|c|}{\textit{Larger k-mers without pretraining}}\\
    \midrule
    $k=10$ & 0.486 & 0.626 &  0.547\\
    $k=12$ & 0.495 & 0.639 & 0.558 \\
    \bottomrule
    \end{tabular}
    }
    \label{tab:ablation}
\end{table}

\textbf{Effect of Structural Priors.} We also evaluate the effect of structural priors on the \textit{Metagenome2Vec} representations by changing how they are constructed. We consider three different settings - no global priors to initialize the embeddings (Section~\ref{sec:globalStructure}), without normalizing the edge weights in the global graph structure (Equation~\ref{eqn:weight_update}) and with only unidirectional attention, i.e., the transformer is trained as an autoregressive model. It can be seen from Table~\ref{tab:ablation} that the design with the highest impact is the normalization of the edge weights in the global structural graph and could be attributed to the fact that the biased random walks for capturing the global structure would be highly impacted by spurious patterns that can occur due to observation noise. All alternative design choices have significantly lower recall, which indicates that the use of global priors and bidirectional attention help capture differences between similar reads.

\textbf{Effect of K-Mers}. Finally, we vary the length of each k-mer to identify the optimal structure. Note that we only consider $k{=}3-6$ for the entire pipeline due to scalability issues since $k{>}8$ leads to a rather large vocabulary (more than a million) and hence does not converge. We consider larger k-mers ($k{=}10,12$) with only the global priors and summarize the results in  Table~\ref{tab:ablation}, where it can be seen that $k{=}4$ provides the best results, corroborating evidence from studies showing similarity in tetranucleotide bases between closely related species~\cite{perry2010distinguishing}. 

\section{Discussion and Future Directions}
In this work, we presented \textit{Metagenome2Vec}, a deep neural representation that explicitly captures the global structural properties of the metagenome sequence and produces \textit{contextualized} representations of k-mers. With extensive experiments, we evaluate and demonstrate the effectiveness of using such representations for both targeted and generalized pathogen detection for scalable metagenome analysis. We show that the proposed representation offers competitive performance to traditional bioinformatics pipelines for segmenting metagenome samples into genome reads from different pathogen species with limited labeled data. 
We draw inspiration from these results and aim to integrate deep neural representations into traditional bioinformatics pipelines to accelerate point-of-care diagnosis.
\section{Acknowledgement}
This research was supported in part by the US Department of Agriculture (USDA) grants AP20VSD and B000C011.

We thank Dr. Kitty Cardwell and Dr. Andres Espindola (Institute of Biosecurity and Microbial Forensics, Oklahoma State University) for providing access and assisting with use of the MiFi platform. 

\bibliography{egbib}
\bibliographystyle{IEEEtran}

\end{document}